\documentclass[twocolumn,showpacs,preprintnumbers,amsmath,amssymb]{revtex4}
\usepackage{graphicx}% Include figure files
\usepackage{dcolumn}% Align table columns on decimal point
\usepackage{bm}% bold math
\begin{document}
\preprint{APS/123-QED}
\title{THE METHOD OF THE DESCRIPTION OF
DYNAMICS NONEQUILIBRIUM SYSTEMS WITHIN THE FRAMES OF THE CLASSICAL
MECHANICS}
% Force line breaks with
\author{V.M. Somsikov}
 \altaffiliation[] {}
 \email{nes@kaznet.kz}
\affiliation{%
Laboratory of Physics of the geoheliocosmic relation, Institute of
Ionosphere, Almaty, Kazakstan.
}%

\date{\today}% It is always \today, today,
\begin{abstract}
Within the frames of the analytical mechanics the method of the
description of dynamics of nonequilibrium systems of potentially
interacting elements is develops. The method is based on an
opportunity of representation of nonequilibrium system by set of
interacting equilibrium subsystems. The equation of motion of
interacting subsystems is found. Based on it the Lagrange, Hamilton
and Liouville equations for subsystems are obtained. The expression
of an entropy production is found. The way of a substantiation of
thermodynamics in the frames of classical mechanic is offered.
\end{abstract}

%Use showkeys class option if keyword
\maketitle

\section{\label{sec:level1}Introduction\protect}

The investigations of the open nonequilibrium systems collided with
the big difficulties. One of them is contradiction between classical
mechanics and thermodynamics. Since Boltzmann and up to now the
attempts to remove this contradiction are undertaken [1-4]. But, as
a rule, offered solution cannot overcome Poincare's theorem of the
recurrence which forbid an establishment of equilibrium in
Hamilton's systems [2, 3]. The contemporary explanation of the
mechanism of an establishment of equilibrium in Hamilton's systems
basing on the condition of mixing is not rigorous enough as it
demands postulation "coarse grain", i.e. averaging of the phase
space on physically small volume. The nature of such average is
impossible to explain in the frame of the classical mechanics.

With the purpose of finding-out of the physical reason of the
contradiction between thermodynamics and classical mechanics, we had
been analyzed a hard-disks system [5, 6]. The non-equilibrium system
of disks was represented by a set of interacting equilibrium
subsystems (IES). Their analysis has shown that the process of an
establishment of equilibrium is caused by transformation of kinetic
energy of IES motion to their internal energy in a result of a chaos
increasing of the disks velocities. I.e. the dissipation for a disks
system is determined by the transformation of energy of relative
motion of IES into internal energy by the work of the collective
force of interaction IES. Therefore the problem about evolution of
nonequilibrium systems is reduced to determining of the forces
between IES. For a hard-disks system of these forces are easily
enough determined due to the energy of disks has only a kinetic
component. But fundamental forces of interactions is potential,
therefore the establishment of equilibrium is determined by the
forces which work redistributes between IES both kinetic, and
potential energy. Therefore to determine the process of evolution in
system, it is necessary to determine, what character energy streams
between IES created by the intrasystem forces. It is impossible to
do basing on the Hamilton formalism because formalism is applicable
only for system as a whole. However streams of energy between IES
can be determined with the help of equation of the systems (EQS)
which determine energy exchange between IES [7].

The EQS allows offering the new approach to the analysis of
nonequilibrium systems. This approach is based on the following
assumptions and conditions: 1). The closed nonequilibrium system can
be submitted by a set of motioning in relative to each other of
interacting IES; 2). Dynamics of nonequilibrium system is equivalent
of the dynamics of interacting IES; 3). The energy of IES should be
submitted by the sum of internal energy and energy of IES motion as
the whole; 4). Each element of system should be fixed for
corresponding IES without dependence from its mixing in space; 5).
During all process of a subsystem are considered as equilibrium.

The applicability of the first condition is guaranteed by that it
was successfully used earlier, for example, in case of a
substantiation of a principle of entropy maximum for the
equilibrium systems [8]. The second condition follows from the
first. The third condition is necessary for introduction in the
description of dynamics of systems of internal energy, as one of
key parameters describing dynamics of interacting IES. The fourth
condition allows avoiding a problem of redefinition IES due to
particles mixing. Last condition is known from statistical
physics. It, not depriving a task of a generality, removes the
problems connected to complication of the analysis of
redistribution of various types of energy between IES. The offered
approach allows to connect thermodynamics with classical mechanics
and to analyze some laws of evolution of nonequilibrium system.

Here the substantiation of this approach is offered. The EQS is
obtained. The expression for the dissipative force determining
change of internal energy IES is obtained also. The analysis of
Lagrange, Hamilton and Liouville equations for IES is given.
Formulas for the entropy and it production are submitted. The
connection EQS and the basic equation of thermodynamics is shown.

\section{THE EQUATION OF SYSTEM DYNAMICS}

The analytical mechanics as usually constructed from a postulate:
"virtual work of forces of reaction is always equal to zero on any
virtual displacement which is not breaking set kinematics
connections" [9]. Basing on this postulate and a condition of
monogenic of the active forces, come to a principle of the least
action, functions Lagrange and Hamilton. It is possible to build
mechanics based on the principle of the least action [10]. But it is
possible to do basing on the law of conservation of energy without
the requirement monogenic forces. Here by this way the equation of
the systems interaction will be obtained. It will help us discover
that the dynamics of the system is determined by two types of
energy: energy of motion of system as the whole and internal energy.
To each of this type of energy the force which change this type are
corresponds.

Let us obtain the dynamics equations for elements and their
systems in a field of external forces. If time is homogeneous, for
energy of system, $E$, we have: $\dot{E}=0$. In non-homogeneous
space at presence of interaction of particles, energy of system
depends on both velocities of elements, $v$, and from their
coordinates, $r$. In this case the equality, $\dot{E}=0$, takes
place only when the energy depends on two additive parts. One part
should be a function of velocity, and the second one -
coordinates. I.e. the energy can be written down as:
$E=\varphi{[T+U]}=const$, where $T=\sum\limits_{i=1}^{N}
T_i({v_i}^2)$, $v_i$ is a velocity of $i$- element, $T$ is the
kinetic energy of the system, and $U(r)$ is the potential one
[10].

The $\varphi$-function should be linear in order to be constant when
the coordinates and velocities are being changed. It is always
possible to represent such function as $E=T+U$ by means of scale
transformation and usage of the necessary system of coordinates.
Thus, the sum of kinetic and potential energies of the system in a
non-homogeneous space should be constant.

Let us take elementary particle with mass $m$ and velocity $v$. The
kinetic energy corresponding to the particle will be
$T(v^2)=mv^2/2$, and the potential energy -$U(r)$, so
$E=m{v^2}/2+U(r)=const$. In this case from equality $\dot{E}=0$
follows that:
\begin{equation}
v(m\dot{v}+\partial{U}/\partial{r})=0\label{eqn1}
\end{equation}
The eq. (1) is a balance equation of the kinetic and potential
energies. The first term of the equation determines the change of
energy caused by inertness of a particle, and the second term is
the change of energy caused by heterogeneity of space. It is
obvious, that the equation (1) is carried out, if the condition
takes place:
\begin{equation}
m\dot{v}=-\partial{U}/\partial{r}\label{eqn2}
\end{equation}
It is Newton equation (NE). This equation is determining the
connection of acceleration of the particle with the external force.
The right hand side of eq. (2) is the active force. The left hand
side is the inertial force [9, 10]. The particle moves along the
gradient of a potential function. The work of forces on the closed
line in a potential field is equal to zero. The dynamics of the
particle is reversible.

Let us take a system, which consists of $N$ potentially
interacting elements; the mass of each element is equal to 1.  The
force acting on each element is equal to the sum of the forces
from another elements and the external force. The force between
elements is central and depends on the distance between them.

The energy of the system is equal to the sum of kinetic energies
of elements - $T_N=\sum\limits_{i=1}^{N} m{v_i}^2/2$, potential
energies in the field of external forces - ${U_N}^{env}$, and the
potential energy of their interaction
${U_N}(r_{ij})={\sum\limits_{i=1}^{N-1}}{\sum\limits_{j=i+1}^{N}}U_{ij}(r_{ij})
$, where $r_{ij}=r_i-r_j$ - is a distance between elements $i$ and
$j$. So, $E=E_N+U^{env}=T_N+U_N+U^{env}=const$.

The time derivative of the energy will be as follows:

\begin{equation}
{\sum\limits_{i=1}^{N}}v_i\tilde{F}_{i}=0 \label{eqn3}
\end{equation}

Where $\tilde{F}_i=m\dot{v}_i+
\sum\limits_{j\neq{i}}^{N}F_{ij}+F_i^{env}$ is effective force for
$i$ particle; $\dot{U}_{env}=\sum\limits_{i=1}^{N}v_iF_i^{env}$;
$F_{ij}=\partial{U_N}/\partial{r_{ij}}$;
$F_{i}^{env}(r_i)=\partial{U^{env}}/\partial{r_{i}}$.

The eq. (3) can be rewritten as:
${\dot{E}=\sum\limits_{i=1}^{N}}v_i\tilde{F}_{i}=0$. This equality
can be treated as orthogonality of the vector of effective forces
with respect to the vector of velocities of elements of the
system. If there are no restrictions imposed on the $v_i$
directions, the requirement $\tilde{F}_i=0$ is satisfied [9]. Then
from eq. (3) we obtain:

\begin{equation}
{m\dot{v}_i=-\sum\limits_{i=1}^{N}}v_i\tilde{F}_{i}-F_i^{env}
\label{eqn4}
\end{equation}

It is NE for the system's elements in non-homogeneous space. From it
follows, that the motion of an element of system is determined by
the force equal to the sum of vectors of forces, acting from all
other particles and external force [11, 12].

Let us obtain an equation of motion of the system as a whole in an
external field. In variables in which energy of system is
represented in the form of the sum of energy of its motion and the
internal energy, the external force acting on system also is
represented us a two corresponding forces. For it show, we shall
take advantage of equality: $T_N=\sum\limits_{i=1}^{N}
m{v_i}^2/2=m/(2N)\{V_N^2+\sum\limits_{i=1}^{N-1}\sum\limits_{j=i+1}^{N}v_{ij}^2\}$
(a), where $V_N=\dot{R}_N=1/N\sum\limits_{i=1}^{N}\dot{r}_i$ -are
velocities of the center of mass (CM); $R_N$ - are coordinates of
the CM; $v_{ij}=\dot{r}_{ij}$. We will write the energy of the
system in such a way: $E_N=T_N^{tr}+E_N^{ins}$,
$E_N^{ins}=T_N^{ins}+U_N$. Then the eq. (3) can be written as
follows:
\begin{equation}
\dot{T}_N^{tr}+ \dot{E}_N^{ins}
=-\sum\limits_{i=1}^{N}v_iF_i^{env}\label{eqn5}
\end{equation}
Where $\dot{T}_N^{tr}=M_NV_N\dot{V}_N; M_N=mN$;
$\dot{E}_N^{ins}=\dot{T}_N^{ins}+\dot{U}_N^{ins}$=
$\sum\limits_{i=1}^{N-1}\sum\limits_{j=i+1}^{N}v_{ij}(m\dot{v}_{ij}/N+F_{ij})$.

Let us represent velocity of the motion of elements of the system as
the sum of velocities of their motion with respect to the CM of the
system - $\tilde{v}_i$, and velocity of the CM of itself - $V_N$ ,
i.e. $v_i=\tilde{v}_i+V_N$. Using these variables we will have the
following: $T_N=\sum\limits_{i=1}^{N}
m{v_i}^2/2=m/(2N)V_N^2+mV_N\sum\limits_{i=1}^{N}\tilde{v}_i
+\sum\limits_{i=1}^{N}m\tilde{v}_{i}^2/2$. As
$\sum\limits_{i=1}^{N}\tilde{v}_i=0$, then
$T_N=m/(2N)V_N^2+\sum\limits_{i=1}^{N}m\tilde{v}_{i}^2/2$. Therefore
$\sum\limits_{i=1}^{N}m\tilde{v}_{i}^2/2=
1/(2N)\sum\limits_{i=1}^{N-1}\sum\limits_{j=i+1}^{N}v_{ij}^2$. Thus
the kinetic energy of the relative motion of particles of the system
equals the sum of kinetic energies of the particles' motion with
respect to the CM.

Let us take into account that
$r_{ij}=\tilde{r}_{ij}=\tilde{r}_i-\tilde{r}_j$, where
$\tilde{r}_i, \tilde{r}_j$ - are coordinates of the elements with
respect to the system's CM. In this case we can write:
$U_N(r_{ij})=U_N(\tilde{r}_{ij})=U_N(\tilde{r}_i)$. That is
$\sum\limits_{i=1}^{N-1}\sum\limits_{j=i+1}^{N}v_{ij}F_{ij}(r_{ij})
=\sum\limits_{i=1}^{N}\tilde{v}_iF_i(\tilde{r}_i)$, where
$F_i=\partial{U_N}/\partial{\tilde{r}_i}
=\sum\limits_{j\neq{i},j=1}^{N}\partial{U_N}/\partial{r_{ij}}$. By
means of generalization of corresponding equalities for the
kinetic and potential energies of the system, we will obtain from
eq. (5):
\begin{eqnarray}
V_NM_N\dot{V}_N+
\sum\limits_{i=1}^{N}m\tilde{v}_i(\dot{\tilde{v}}_i+F(\tilde{r})_i)=\nonumber\\=
-V_NF^{env}-\sum\limits_{i=1}^{N}\tilde{v}_iF_i^{env}(R,\tilde{r}_i)\label{eqn6}
\end{eqnarray}

Here $F^{env}=\sum\limits_{i=1}^{N}F_i^{env}(R,\tilde{r}_i)$, $R$
- is a coordinate of CM.

The eq. (6) determines the balance of energy of the system in
non-homogeneous space. The first term in the right hand side
corresponds to change of kinetic energy of motion of system as the
whole. The second term determines the change of internal energy.
If the external forces are absent the eq. (6) breaks up on
independent equations: one of them is the equation of motion of
the CM, and the others are the equations of motion of particles.

Let us take into account that $F^{env}=F^{env}(R+\tilde{r}_i)$,
and suppose that $R\gg\tilde{r}_i$. Then it is possible to expand
the force $F^{env}$ in a series using a small parameter,
$\tilde{r}_i/R$. Keeping the terms up to first-order of
infinitesimal, we will have:
$F_i^{env}=F_i^{env}|_{R}+(\nabla{F_i^{env}})|_{R}\tilde{r}_i\equiv
F_{i0}^{env}+(\nabla{F_{i0}^{env}})\tilde{r}_i$. Taking into
account that $\sum\limits_{i=1}^{N}\tilde{v}_i
=\sum\limits_{i=1}^{N}\tilde{r}_i=0$ and
$\sum\limits_{i=1}^{N}F_{i0}^{env}=NF_{i0}^{env}=F_0^{env}$, we
can set from (6):
\begin{eqnarray}
V_N(M_N\dot{V}_N)+
\sum\limits_{i=1}^{N}m\tilde{v}_i(\dot{\tilde{v}}_i+F(\tilde{r})_i)\approx\nonumber\\\approx
-V_NF_0^{env}-({\nabla}F^{env}_{i0})\sum\limits_{i=1}^{N}\tilde{v}_i\tilde{r}_i\label{eqn7}
\end{eqnarray}

In the eq. (7) the force  $F_0^{env}$ is potential and depends on
$R$. It is determines the change of kinetic energy of system as the
whole. The second term in the right hand side is depending on
coordinates of particles and their velocities in relative to the CM.
It is determines the change of internal energy of system. The force
corresponding to this term is non-potential and do not change the
system's momentum as the whole.

Thus, dynamics of system in an external field is determined by
transformation of two types of energy: the energy of motion of
system as the whole and its internal energy. For each of this type
of energy the force determining its change is corresponds. Energy of
motion of system is determined by macroparameters - velocities and
coordinates of the CM. The change of internal energy is determined
by macroparameters and microparameters. Thus, the system is similar
to the structured particle. Its dynamics is determined by the forces
changing its internal energy and energy of its motion as the whole.

Let us note, that the eq. (5) can be obtained, directly basing on
the NE for elements. For this purpose we shall multiply the eq.
(4) on the corresponding velocity. After summation the obtained
equations for all particles we shall have the eq. (5) (if we have
summarized the eq. (4) without multiplying it on velocity in this
case the internal forces in the second term of the eq. (5) will be
lost [13]). It is confirms validity of the equation (5).

\section{The equation of interaction of two subsystems}

The nonequilibrium system can be submitted by a set of IES [8].
The dynamics of a set of IES is determined by the eq. (6) if the
external forces replace on the forces of interaction of IES. Thus,
to find the equation of motion of IES, it is necessary to
determine collective forces of their interaction.

Let us the system consists of two IES - $L$ and $K$. We take all
elements to be identical and have the same weight 1, and $L$ to be a
number of elements in $L$ - ILS, $K$ -is a number of elements in $K$
-ILS, i.e. $L+K=N$, $V_L=1/L\sum\limits_{i=1}^{L}v_i$ and
$V_K=1/K\sum\limits_{i=1}^{K}v_i$ - are ILS's velocities with
respect to the CM of system. The velocity of the system's CM we take
equal to zero, i.e. $LV_L+KV_K=0$.

We can represent the energy of the system as
$E_N=E_L+E_K+U^{int}=const$, where $E_L$ and $E_K$ are the ILS, and
$U^{int}$ - is the energy of their interaction. According to the eq.
(6), the energy of each ILS can be represented as
$E_L=T_L^{tr}+E_L^{ins}$, $E_K=T_K^{tr}+E_K^{ins}$, where
$T_L^{tr}={M_L}V_L^2/2$, $T_K^{tr}={M_K}V_K^2/2$, $M_L=mL, M_K=mK$.
$E^{ins}$- is the internal energy of a ILS. The $E^{ins}$ consists
of the kinetic energy of motion of the elements with respect to the
CM of IES - $T^{ins}$ and their potential energy - $U^{ins}$, i.e.
$E^{ins}=T^{ins}+U^{ins}$, where
$U_L^{ins}=\sum\limits_{i_L=1}^{L-1}\sum\limits_{j_L=i_L+1}^{L}U_{{i_L}{j_L}}(r_{i_Lj_L})$,
$U_K^{ins}=\sum\limits_{i_K=1}^{K-1}\sum\limits_{j_K=i_K+1}^{K}U_{{i_K}{j_K}}(r_{i_Kj_K})$.
The energy $U^{int}$ is determined as
$U^{int}=\sum\limits_{j_K=1}^{K}\sum\limits_{j_L=1}^{L}U_{j_Lj_K}(r_{j_Lj_K})$.
Indexes $j_k,j_L,i_K,i_L$ determine belonging of the elements to
corresponding ILS. In equilibrium we have: $T^{tr}=0$. Hence, if the
system aspirates to equilibrium, then $T^{tr}$ energy for each ILS
will be transformed into the internal energy of IES.

We have obtained the equations of dynamics of $L$ and $K$ of IES
in the following way. Let us differentiate energy of system on
time. In order to find the equation for $L$ - IES, at the left
hand side of obtained equality we have kept only those terms which
determine the change of the kinetic and potential energies of
interaction of elements of $L$ - IES. We replaced all other terms
in the right hand side and combined the groups of terms in such a
way when each group contained of the terms with identical
velocities. In accordance with NE (see eq. (5), the groups which
contain terms with velocities of the elements from $K$- IES are
equal to zero. As a result the right hand side of the equation
will contain only the terms which determine the interaction of the
elements  $L$-IES with the elements $K$-IES. The equation for
$K$-IES can be obtains in the same way. Then we execute
replacement of variables: $ v_i=\tilde{v}_i+V$ and take into
account equality (a). As a result we will have [7, 14]:
\begin{eqnarray}
V_LM_L\dot{V}_L+{\sum\limits_{i_L=1}^{L-1}}\sum\limits_{j_L=i_L+1}^{L}\{v_{i_Lj_L}
[\frac{{m\dot{v}}_{i_Lj_L}}{L}+\nonumber\\+F_{i_Lj_L}]\}=-{\Phi}_L-V_L{\Psi}
\end{eqnarray}
\begin{eqnarray}
V_KM_K\dot{V}_K+{\sum\limits_{i_K=1}^{K-1}}\sum\limits_{j_K=i_K+1}^{K}\{v_{i_Kj_K}
[\frac{{m\dot{v}}_{i_Kj_K}}{K}+\nonumber\\+F_{i_Kj_K}]\}={\Phi}_K+V_K{\Psi}
\end{eqnarray}

Here $R_K=(1/K)\sum\limits_{{i_K}=1}^Kr_{i_K}$;
$R_L=(1/L)\sum\limits_{{i_L}=1}^Lr_{i_L}$;
$\Psi=\sum\limits_{{i_L}=1}^LF^K_{i_L}$;
${\Phi}_L=\sum\limits_{{i_L}=1}^L\tilde{v}_{i_L}F^K_{i_L}$;
${\Phi}_K=\sum\limits_{{i_K}=1}^K\tilde{v}_{i_K}F^L_{i_K}$. The
terms: $F^K_{i_L}(R_K,r_{i_L})=\sum\limits_{{j_K}=1}^KF_{i_Lj_K}$
and $F^L_{j_K}(R_L,r_{i_K})=\sum\limits_{{i_L}=1}^LF_{i_Lj_K}$ -
are forces between the corresponding particle of one IES and all
particles of the other IES. The work of these forces determines
the change of energy of IES.

The eqs. (8, 9) are the EQS. They are describe energy exchange
between IES. Independent variables EQS are macroparameters -
coordinates and velocities of motion IES, and also microparameters
- coordinates and velocities of elements. So, EQS connects among
themselves two types of the description: at a macrolevel and at a
microlevel. The description at a macrolevel determines dynamics
IES as the whole, and at a microlevel determines dynamics of
elements IES.

The force, $\Psi$, determines motion of IES as the whole. This
force is the sum of the potential forces acting on elements of one
IES at the side of another IES.

The forces which determined by the terms ${\Phi}_L$ and
${\Phi}_K$, transformed of the motion energy of IES to internal
energy as a result of chaotic motion of elements of one IES in a
field of the forces of another IES. It is non-potential force
which cannot be expressed as a gradient from any scalar function.
These forces is equivalents to dissipative forces. It can be shown
with the help of some transformations of the eqs. (8, 9). For this
purpose we take into account that
$\sum\limits_{{i_L}=1}^Lm\tilde{v}_{i_l}\dot{\tilde{v}}_{i_L}=
(1/L){\sum\limits_{i_L=1}^{L-1}}\sum\limits_{j_L=i_L+1}^{L}
mv_{{i_L}{j_L}}{\dot{v}}_{{i_L}{j_L}}$,
${\dot{U}}_L=\sum\limits_{{i_L}=1}^LF_{i_L}\tilde{v}_{i_L}=
{\sum\limits_{i_L=1}^{L-1}}\sum\limits_{j_L=i_L+1}^{L}F_{i_Lj_L}{\tilde{v}}_{i_Lj_L}$,
$F_{i_L}=\sum\limits_{{j_L}{{\neq}i_L}}^{L}{\partial}U_L/{\partial}
\tilde{r}_{i_L}$ . Then EQS can be rewritten so:
\begin{equation}
M_L\dot{V}_L=-\Psi-{\alpha}_LV_L \label{eqn10}
\end{equation}
\begin{equation}
M_K\dot{V}_K=-\Psi- {\alpha}_KV_K\label{eqn11}
\end{equation}

where ${\alpha}_{L}=-(\dot{E}^{ins}_{L}+{\Phi}_{L})/V^2_{L}$,
${\alpha}_{K}=-(\dot{E}^{ins}_{K}-{\Phi}_{K})/V^2_{K}$,
${\dot{E}}_L^{ins}=\sum\limits_{{i_L}=1}^L{\tilde{v}}_{i_L}
(m{\dot{\tilde{v}}}_{i_L}+F_{i_L})$,
${\dot{E}}_K^{ins}=\sum\limits_{{i_K}=1}^K{\tilde{v}}_{i_K}
(m{\dot{\tilde{v}}}_{i_K}+F_{i_K})$.

Here "$\alpha_L$", "$\alpha_K$" are coefficients determining
efficiency of transformation of energy of relative motion the IES
into internal energy. They are equivalent to the friction
coefficients. Thus the role of friction is reduced to
redistribution of energy of motion IES between their elements.

If the nonequilibrium system is submitted by a set of IES, the
state of the system can be determined by the point in the phase
space which consists from $6R-1$ of independent variables, where
$R$ is a number of IES. In this space the role of a elementary
particle is carried out by the IES. We will call this space as
$S$-space to distinguish it from usual phase space with $6N-1$ of
independent variables. We see that the $S$-space, unlike usual
phase space, is compressed, though total energy of all elements is
preserved. The rate of compression of $S$-space is determined by
velocity of transformation of energy of relative motion the IES
into their internal energy. The volume of compression of $S$-space
is determined by energy of the IES motion.

Thus, EQS determines the IES dynamics as a result of
transformation of IES interaction energy into two types of energy:
internal energy of IES and energy of its motion. The forces which
are carrying out such transformation break up on potential and
non-potential parts. Potential forces determine the change of the
velocity of IES. The non-potential force determines the change of
their internal energy. The work of non-potential forces connected
with the chaotic motion of elements for one IES in a field of
forces of another IES.

\section{The Lagrange, Hamilton and Liouville equations for IES}

Let us explain how Lagrange, Hamilton and Liouville equations for
the nonequilibrium system which submitted by a set of IES can be
obtained [14]. The canonical types of these equations for elements
are following from the integral principle of Hamilton [9]. In turn
the integral principle of Hamilton follows from differential
principle of D'Alambert. D'Alambert equation is constructed on the
basis of NE. In accordance with D'Alambert principle: "the virtual
work of the effective forces which includes the inertial and
active forces is equal to zero for all reversible virtual
displacements of elements compatible with the given restrictions"
[9]. But if the system consist of a set of IES this principle will
be as follows: the sum of works of all forces of interaction the
IES at their virtual displacements compatible to restrictions on
dynamics is equal to zero.

If the change of the internal energy of the IES can be neglected,
the work on their motion will be determined only by the potential
part of interaction forces of IES. In this case instead of EQS we
can use NE and accept the IES as elementary particle. As a result
we come to the well known canonical equations of a classical
mechanics.

If neglecting by the change of internal energy is impossible,
D'Alambert equation should be written down on the basis of EQS
which take into account the transformation of the IES motion
energy into internal energies in the result of the work of the
non-potential part of forces. As a result we will obtain Lagrange,
Hamilton and Liouville equations for IES. Let us explain briefly a
way of these equations of obtaining. Firstly, the D'Alambert
equation on the basis of EQS  can be obtained. Basing on it we
obtain Lagrange equation. After that the Hamilton and Liouville
equations are deduced.

Let us take a system consisting of $N$ elements which can be
represented by a set of IES. Required Lagrange equation for
nonequilibrium system looks like [14]:
\begin{equation}
d/dt({\partial}{\Im}/{\partial}v_n)-{\partial}{\Im}/{\partial}r_n=-F_n
\label{eqn12}
\end{equation}

Here $\Im$- is a Lagrange function for the system, $n=1,2,3...N$
-is a number of the particle.

At absence of internal degrees of freedom in IES, the work of
non-potential forces is equal to zero. In this case the right hand
side of eq. (3.1) is equal to zero and the equation becomes
canonical [10-12].

The Lagrange equations for two $L$ and $K$ of IES can be written
as:

\begin{equation}
d/dt({\partial}{\Im}_L/{\partial}v_{i_L})-{\partial}{\Im}_L/{\partial}r_{i_L}=-F_{i_L}^K
\label{eqn13}
\end{equation}

\begin{equation}
d/dt({\partial}{\Im}_K/{\partial}v_{i_K})-{\partial}{\Im}_K/{\partial}r_{i_K}=-F_{i_K}^L
\label{eqn14}
\end{equation}
Here ${\Im}_L$ and ${\Im}_K$ - are Lagrange functions for the IES.

The Hamilton's equations for non-equilibrium systems can be
written as:
\begin{equation}
{\partial}{H}/{\partial}r_{n}=-\dot{p}_n-F_n
 \label{eqn15}
\end{equation}

\begin{equation}
{\partial}{H}/{\partial}p_{n}=v_n \label{eqn16}
\end{equation}

The non-potential part of force of interaction IES together with
potential force determines the right hand side of the eq. (15).

The Liouville equation for non-equilibrium systems can be written
as [6, 7]:

\begin{equation}
df/dt=f{\partial}F_n/{\partial}p_n \label{eqn17}
\end{equation}
Her $f$-is a distribution function for the system's particles.

The right hand side of eq. (17) is not equal to zero as forces
between IES depend on velocities. It means that $S$-space for the
nonequilibrium case is compressed. When relative motions of IES will
disappear, the right hand side of eq. (17) will be equal to zero.
Thus the descriptions in $S$-space and in the usual phase space for
equilibrium systems are similar. I.e., at absence of IES motion
energy, the dissipative processes do not exist.

\section{EQS and thermodynamics}

Here we would like to show, how basing on the EQS it is possible to
come from the classical mechanics to thermodynamics.

Let us take the motionless nonequilibrium system consisting from
"$R$" of IES. Each of IES consists from enough plenty of elements.
Let us, $dE$ is a work which done above the system. In
thermodynamics it is term as internal energy of a system (do not
confuse $E$ with the $E^{ins}$ - internal energy of IES). The $dE$
is determined by the basic equation of thermodynamics as:
${dE=dQ-PdY}$ [8]. Here, according to common terminology, $E$ is
energy of a system; $Q$ is thermal energy; $P$ is pressure; $Y$ is
volume.

As well as the basic equation of thermodynamics, EQS also is
differential of two types of energy. According to the EQS the
volume $dE$ is redistributed inside of the system so, that one
part of its goes on change of energy of relative motion of IES,
and another part changes their internal energy.

The first term in the left hand side of EQS is a change of kinetic
energy of motion of a IES as the whole, $dT^{tr}$. This term
corresponds to the term ${PdY}$. Really [8],
${dT^{tr}=VdV=V\dot{V}dt=\dot{V}dr=PdY}$.

If the potential energy is a homogeneous function of a second
order of the radius-vectors, then as it follows from the virial
theorem [10], we have:
${\bar{E}^{ins}=2\bar{\tilde{T}}^{ins}=2\bar{\tilde{U}}^{ins}}$.
The line denotes the time average. Let us consider the system near
to equilibrium. The average energy of each element is
${\bar{E}^{ins}={E}^{ins}/N=\kappa{T}_0^{ins}}$ where $N$ is a
number of elements. As the increasing of the internal energy is
determined by the volume ${dQ}$, then we will have:
${dQ\approx{T}_0^{ins}[d{E}^{ins}/{T}_0^{ins}]
\sim{T}_0^{ins}[{dv_0}/{v_0}]}$, where ${v_0}$ is the average
velocity of an element, and ${dv_0}$ is its change. For the system
in the closed volume we have:
${dv_0/v_0\sim{{d\Gamma}/{\Gamma}}}$, where ${\Gamma}$ is the
phase volume of a system, ${d\Gamma}$ will increase due to
increasing of the system's energy on the value, ${dQ}$. By keeping
the terms of the first order we get:
${dQ\approx{T}_0^{ins}d\Gamma/\Gamma={{T}_0^{ins}}d\ln{\Gamma}}$.
By definition ${d\ln{\Gamma}=dS^{ins}}$, where ${S^{ins}}$ is a
entropy [8]. So, near equilibrium we have
$d\bar{E}^{ins}={dQ\approx{T}_0^{ins}dS^{ins}}$.

If we break IES into subsystems, these subsystems will not have
the relative motion. Therefore the entropy increasing,
$\Delta{S}$, for nonequilibrium system is completely determined by
the energy $T^{tr}$ passing into $E^{ins}$. Therefore $\Delta{S}$
can be determined by the formula [7, 14]:

\begin{equation}
{{\Delta{S}}={\sum\limits_{l=1}^R{\{{N_l}
\sum\limits_{k=1}^{N_l}\int{\sum\limits_s{{\frac{{F^{L}_{ks}}v_k}{E^{L}}}}{dt}}\}}}}\label{eqn18}
\end{equation}

Here ${E^{L}}$ is the kinetic energy of $L$-IES; $N_L$ is the
number of elements in $L$-IES; $L=1,2,3...R$; ${R}$ is the number
of IES; ${s}$ is a number of the external element which
interaction with element ${k}$ belonging to the $L$-IES;
${F_{ks}^{L}}$ is a force, acted on $k$-element; $v_k$ -is a
velocity of the $k$- element.

Thus in agreement with eq. (18), the entropy is determined by the
energy of the relative motion of IES, transformed in an internal
energy as a result of work of non-potential part of forces between
IES.

To obtain equation for the entropy production we take into account
that: $\Delta{S}=\Delta{Q}/T$. Thus
$dS/dt=[dE^{ins}/dt]/(kE^{ins}$, where $k$ is a coefficient. It is
possible to express this formula through the work of forces of
interaction IES. Let us $E_0$ is a full system's energy,
$E^{tr}_{L0}$ is a beginning energy of relative motion of $L$-IES.
In according with eqs. (10,11) the rate of increasing of the
internal system's energy is equal to: $\zeta=\sum^R_{L=1}\Phi_L$.
The internal energy of a system is equal to:
$E^{ins}=E_0-\sum^R_{L=1}E^{tr}_L$, where $\sum^R_{L=1}E^{tr}_L$
is a sum of IES energy of relative motion. But
$\sum^R_{L=1}E^{tr}_L=\zeta_0-\int^t_0\zeta(t){dt}$, where
$\zeta_0=\sum^R_{L=1}E^{tr}_{L0}$. Then entropy production for the
system, $\varrho_{prod}=dS/dt$, can be write:
$\varrho_{prod}=D/(1-D_0+\int^t_0D(t)dt)$, where $D=\zeta/E_0$,
$D_0=\zeta_0/E$.

The energy of $T^{tr}$ characterizes the rate of non-equilibrium
system while $E^{ins}$ characterizes its degree of equilibrium. To
be a stationary state of the system the loss of energy $E^{tr}$ is
necessary to compensate by the inflow of external energy. It is
possible as a result of contact of system with thermostat or due
to outflow of radiation. Then the stationary state of
nonequilibrium system is characterized by the formula:
$\varrho_{prod}=|\varrho_{-}|-|\varrho_{+}|$, where $ \varrho_{-}$
is entropy outflow, $\varrho_{+}$ -is inflow of entropy,
$\varrho_{prod}$ is entropy production, determined by the formula
(18).

Thus, the EQS leads to the consent between classical mechanics and
thermodynamics. It is possible to explain by that the EQS equation
determines dynamics of the structured particles taking into
account the change of their internal energy.

\section{Conclusion}
Creation of methods of studying of dynamics of the open
nonequilibrium systems can be carried to a task: how to determine
the properties and laws of a systems dynamics, knowing properties
and laws of dynamics of its elements, at the external restrictions
set on system. Within the frame of existing formalisms of a
classical mechanics its solution collided with the contradiction
between classical mechanics and thermodynamics. According to
obtained here results the nature of this contradiction consists in
the following.

Evolution is impossible for systems of elementary particles as
processes determining it are caused by the dissipative forces. But
such opportunity appears for the structural particles possessing an
internal energy. The change of internal energy causing evolution is
related to work of non-potential part of the interaction force of
particles. In many cases: near to equilibrium, in the linear
approach of the theory of perturbation, etc., the neglecting of this
part of force is admissible. But, it is obvious, that at studying
evolutionary processes its account is necessary.

Evolution of nonequilibrium system is determined by the
intrasystem energy streams. Our key idea is that these energy
streams can be found with the help of EQS if the system is
submitted by a set of IES. Dynamics of IES is determined by the
internal energy. Only owing to occurrence of this parameter within
the frame of classical mechanics the determination of entropy is
appearing as the rate describing increase of internal energy of
the IES.

Independent variables in EQS are macroparameters - coordinates and
velocities of IES motion and also microparameters - coordinates
and velocities of elements. I.e., EQS connects two types of the
description: at a macrolevel and at a microlevel. This description
is achieved by representation of IES energy by the sum of two
types of energy: energy of their motion and internal energy. The
change of each of them due to the IES interaction are determined
by the corresponding type of collective force between of IES.

The change of the motion energy of IES is determined by the work
of potential forces. The change of the internal energy of IES is
determined by the work of non-potential forces as a result of
chaotic motion of particles. The process of transforming of
internal energy into the energy of IES motion is forbidden by the
law of preservation of momentum. Really, the IES momentum cannot
be changed by internal motion of its elements. Therefore the work
on the closed contour for IES is not equal to zero due to
increasing of the internal energy of IES.

The Lagrange, Hamilton and Liouville equations on the basis of EQS
are deduced. These equations describe dynamics of the
nonequilibrium system in the compressible $S$-space where points
is corresponds to the momentum and coordinates of the IES.

The following mechanism of irreversibility can be proposed: the
energy of relative motion of IES is transformed into their internal
energy as a result of the work of the non-potential part of
interaction force between of IES. The system equilibrates, when
relative motion of IES disappears.

The explanation of the First law of thermodynamics is based on the
fact that the work of subsystems' interaction forces changes both
the energy of their motion and their internal energy. The
explanation of the Second law of thermodynamics is based on the
condition of irreversible transformation of the subsystems' relative
motion energy into their internal energy.

The dynamics of nonequilibrium systems is determined by other form
of symmetry, than dynamics of the equilibrium systems. Really, both
for elementary particles, and for equilibrium systems the dynamics
is reversible because internal energy are absent. But for
nonequilibrium systems these types of symmetry are broken. It is due
to the work of the non-potential forces of IES interaction that
change the internal energy. Thus it is possible to assume, that the
broken of symmetry (for example, observable broken $CP$-symmetry in
disintegrations $\beta$ - meson [15]), can testify that products of
disintegration are the structured particles with time of a life
commensurable with the period of supervision.

If we take into account that all particles are structured we come to
the conclusion about existence of infinite hierarchical sequence of
the microstructures enclosed in macrostructures. In turn, a
macrostructure are a microstructure for the following hierarchical
step. The occurrence and existence of such hierarchical sequence in
the nature is provided with hierarchy of fundamental forces of
interaction. As a rule, the force of interaction of the bottom
hierarchical level there are more than forces of the top level.
Indeed, nuclear forces more than electromagnetic forces and
electromagnetic forces more than the gravity force. We come to the
infinite divisibility of a matter because of impossibility of
occurrence of elementary particles. This conclusion has the
experimental confirmation[15, 16].

Thus the offered approach allows us to expand a scope of formalism
of classical mechanics on the open nonequilibrium systems and
eliminate contradictions between classical mechanics and
thermodynamics.

\smallskip

\end{document}